\pdfoutput=1
\RequirePackage{color}
\documentclass{JINST}

\let\ifpdf\relax

\usepackage{graphicx}

\usepackage{color}

\usepackage[caption=false]{subfig}
\makeatletter
\AtBeginDocument{%
\DeclareRobustCommand*\subref{\@ifstar\sf@@subref\sf@subref}}
\makeatother

\usepackage{lineno}

\usepackage{setspace}

\definecolor{codebgcolor}{rgb}{0.8,0.8,0.8}
\definecolor{darkblue}{rgb}{0.0,0.0,0.3}

\DeclareGraphicsExtensions{.pdf,.png,.jpg}

\title{Liquid Argon Dielectric Breakdown Studies with the MicroBooNE Purification System}

\author{
R.~Acciarri, 
B.~Carls, 
C.~James, 
B. Johnson, 
H.~Jostlein, 
S.~Lockwitz\thanks{Corresponding author: lockwitz@fnal.gov (S.~Lockwitz)},
B.~Lundberg,
J.L.~Raaf, 
R.~Rameika, 
B. Rebel, 
G.P.~Zeller, 
and M.~Zuckerbrot\\
Fermi National Accelerator Laboratory, P.O. Box 500, Batavia, IL, 60510, USA }

\abstract{
The proliferation of liquid argon time projection chamber detectors makes the characterization of the dielectric properties of liquid argon a critical task. To improve understanding of these properties, a systematic study of the breakdown electric field in liquid argon was conducted using a dedicated cryostat connected to the MicroBooNE cryogenic system at Fermilab. An electrode sphere-plate geometry was implemented using spheres with diameters of 1.3~mm, 5.0~mm, and 76~mm. The MicroBooNE cryogenic system allowed measurements to be taken at a variety of electronegative contamination levels ranging from a few parts-per-million to tens of parts-per-trillion. The cathode-anode distance was varied from 0.1~mm to 2.5~cm. The results demonstrate a geometric dependence of the electric field strength at breakdown. This study is the first time that the dependence of the breakdown field on stressed cathode area has been shown for liquid argon.}


\begin{document}

\section{Motivation}

Liquid argon time projection chambers (LArTPCs) are a popular and effective detection technique for neutrino interactions thanks to their capability to perform three-dimensional imaging of charged particle interactions with fine spatial resolution and accurate calorimetry. Neutrino experiments are driving the design of these detectors to contain increasingly larger volumes of argon. The increase in argon volumes implies both the application of higher voltages and the capability to maintain such voltages in the detector, avoiding discharge.

In this context, a precise characterization of the dielectric properties of liquid argon becomes extremely important.  Several research activities have been devoted to this subject~\cite{ref:swan,ref:swan2,ref:bern, ref:bay}. Nonetheless, parameters related to dielectric breakdown through liquid argon are still not well known due to the varied values found among studies characterized by different liquid argon electronegative contamination levels, electrode sizes, electrode shapes, and materials. 

To address the origin of these differences, a systematic study was conducted of the breakdown electric field in liquid argon as a function of electrode size, liquid argon electronegative contamination level, and cathode-anode distance. These studies were conducted at Fermilab's Liquid Argon Test Facility using a dedicated test stand linked to the MicroBooNE cryogenic system.

\section{Experimental Setup}\label{sec:setup}

\subsection{HV breakdown test device}\label{subsec:HVC}
\begin{figure}[htb]
\centering
\includegraphics[width=0.6\textwidth]{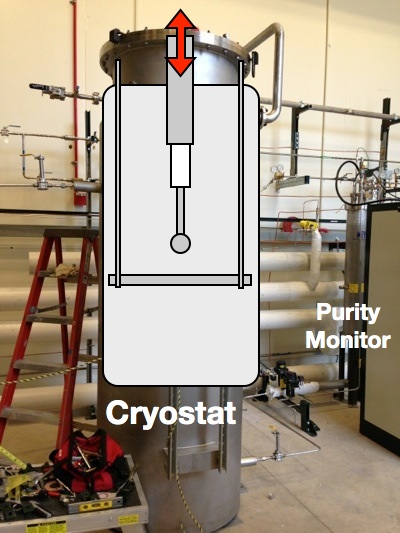}
\caption{Schematic overlaid on a photo of the test stand with spherical-tipped high-voltage feedthrough and adjustable distance between sphere and grounded anode plate.}
\label{fig:expt_setup}
\end{figure}

\begin{figure}[htb]
\centering
\includegraphics[width=0.75\textwidth]{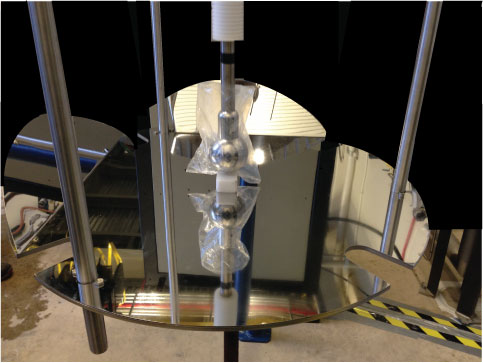}
\caption{Breakdown test device. The stainless steel ground plate was mechanically polished and then electropolished. Angled, mirrored wings were installed for viewing from above. The spherical-tipped HV feedthrough was brought within proximity of the plate for the tests. The white block seen in the photograph was used to calibrate the device and was removed during operation.}
\label{fig:mirror}
\end{figure}

The experimental setup shown in Fig.~\ref{fig:expt_setup} consists of a breakdown test device housed in a cryostat that was connected to the MicroBooNE liquid argon filtration and circulation system. The breakdown measurements reported here were made using a sphere-plate geometry as shown in the figure. The plate was held at ground and the sphere, attached to the end of a high-voltage (HV) feedthrough, was brought to the desired voltage. The 74~cm diameter grounded plate was mechanically polished to a mirror finish, then electropolished. Two mirrored, angled wings were attached to the edges of the grounded plate, as seen in Fig.~\ref{fig:mirror}. The wings allow inspection of the sphere-plane gap through viewports on the top of the cryostat. Both the size of the sphere and its distance from the plate were varied in the tests. 

A Glassman LX150N12 power supply capable of delivering up to $-150$~kV was used for these measurements~\cite{ref:glassman}. It was equipped with a Glassman-supplied microcontroller which allowed control by a {\sc{LabView}} program~\cite{ref:labview}. The output of the power supply was connected to a resistive filter pot that served to isolate the stored energy in the setup. The output of the filter pot was then connected to a high-voltage feedthrough with swappable spherical probe tips. 

\begin{figure}[htb]
  \centering
  \subfloat[]{\includegraphics[width=\textwidth]{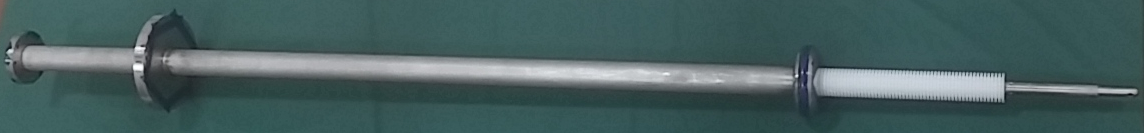}}\\
  \subfloat[]{\includegraphics[width= \textwidth]{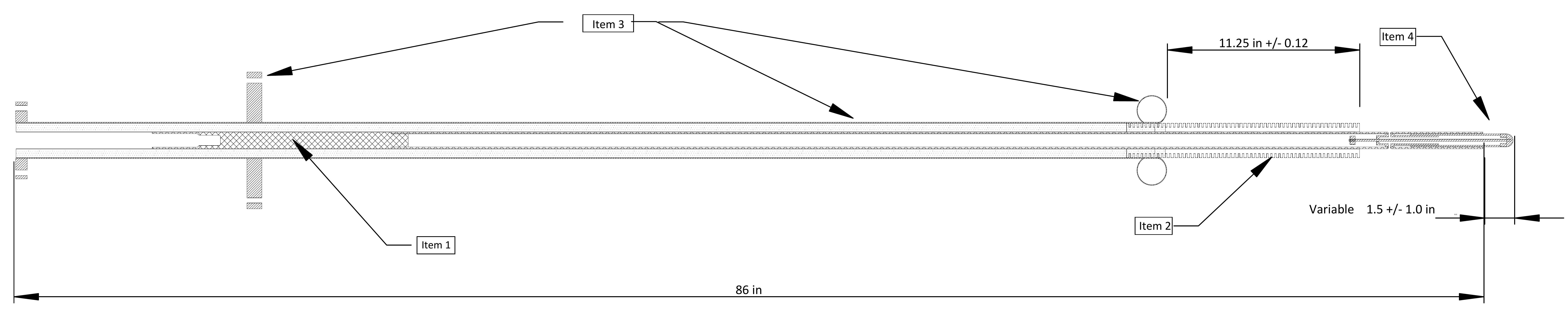}}
  \caption{Photograph (a) and drawing (b) of the production HV feedthrough. The spherical probe tip attached to the end of the inner conductor, on the right side of these figures.}
  \label{fig:ftPict}
\end{figure}

\begin{figure}[htb]
\centering
\includegraphics[width=0.5\textwidth]{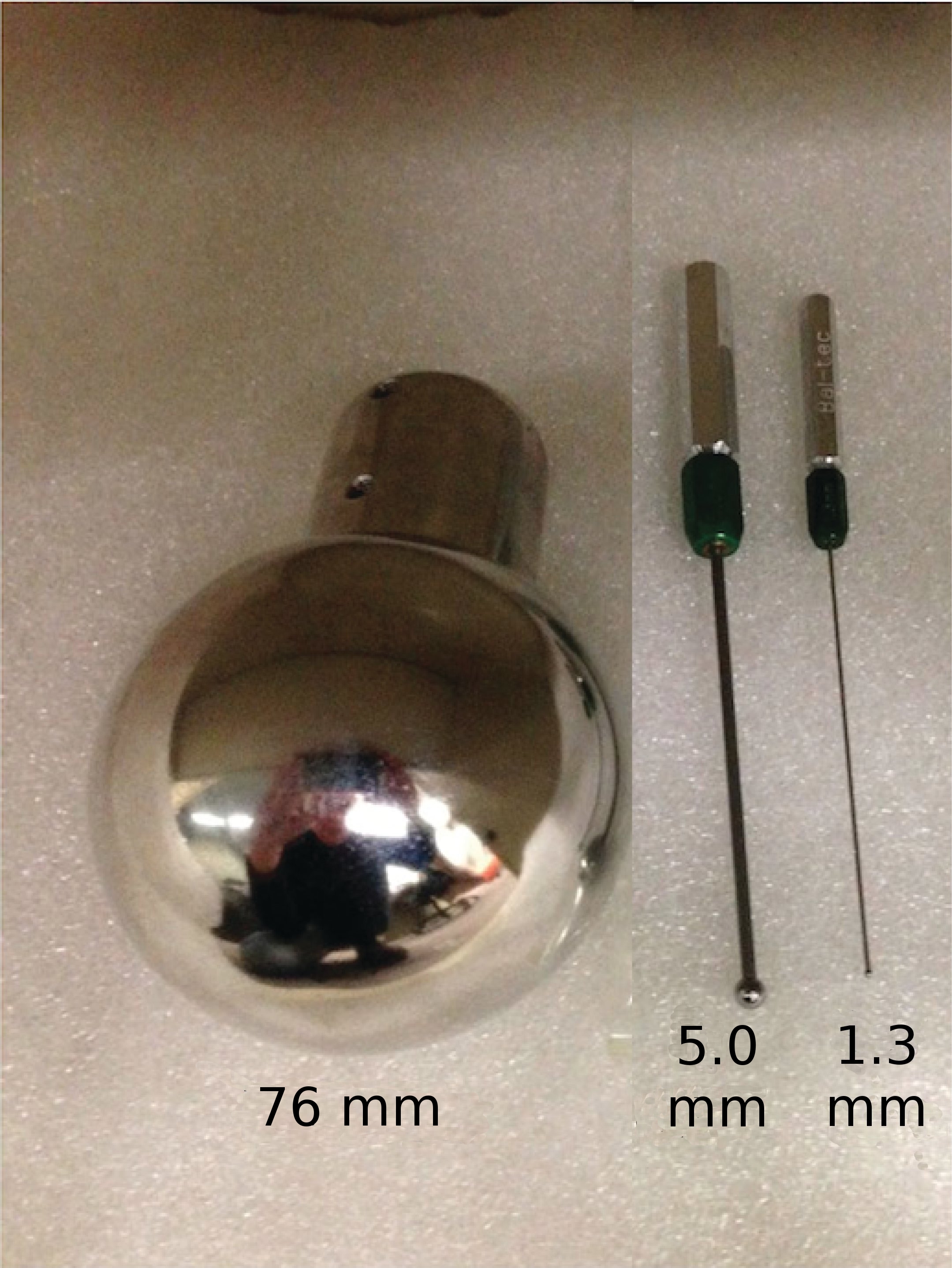}
\caption{Three spherical probe tips used in the testing. The two smaller tips had a chrome finish, while the larger was mirror-finished stainless steel.}
\label{fig:probes}
\end{figure}

The feedthrough was a prototype of the MicroBooNE design, shown in Fig.~\ref{fig:ftPict}. The design was based on the ICARUS feedthrough~\cite{ref:icarus}, with a stainless steel center conductor surrounded by insulating ultra high molecular weight polyethylene (UHMW PE) that was inserted into an outer ground tube. The swappable spherical probe tips attached directly to the center conductor of the feedthrough. The tests reported in this paper evaluated 1.3~mm, 5.0~mm, and 76~mm diameter spherical probe tips, which are shown in Fig.~\ref{fig:probes}.

The feedthrough was attached to a motorized linear translator on the cryostat. The translator was fit with a bellows that allowed the feedthrough to move vertically up and down. 
An OMEGA force transducer model LC703-150~\cite{ref:omega} was connected between the grounded plate and its supports. Force registered by the transducer could then be used to determine contact between the probe tip attached to the feedthrough and the grounded plate.
The vertical distance travelled was read out by an AccuRemote digital readout \#35-812 with 0.01~mm resolution~\cite{ref:accuremote}.

\subsection{Liquid argon filtration and recirculation system}

The cryostat was a vacuum-jacketed vessel held between 0-8~psig or 101-157~kPa for the measurements. The pressure was held at a constant value during each measurement by allowing the argon boil off to vent through a check valve. The pressure varied slightly from one measurement to another. The cryostat was plumbed into MicroBooNE's cryogenic system for access to pure argon. 
The MicroBooNE cryogenic system consists of two filter vessels to remove water and oxygen from the argon, a pump to circulate the argon, and a dewar refrigerated by liquid nitrogen. The first filter was filled with a 4A molecular sieve supplied by Sigma-Aldrich to remove water~\cite{ref:sigmaaldrich}. The second filter was filled with BASF CU-0226~S, a dispersed copper oxide impregnated on a high surface area alumina to remove oxygen~\cite{ref:basf}. The filtration system consistently delivered better than 100~parts-per-trillion (ppt) oxygen equivalent contamination to the test vessel. The contamination level in the vessel was adjusted by adding higher contamination argon from a separate supply dewar. This allowed the production of oxygen equivalent contaminations ranging from 1.5~parts-per-million (ppm) to less than 100~ppt for the testing.

The level of oxygen contamination in the argon was measured using two Servomex gas analyzers, a DF-310E and a DF-560E~\cite{ref:spectris}. Combined, the two oxygen analyzers covered a range of 0.1~parts-per-billion (ppb) to 5000~ppm.  The level of nitrogen contamination in the argon was measured using an LD8000 Trace Impurity Analyzer~\cite{ref:ldetek}.

Additionally, contamination values ranging between 300 and 50~ppt oxygen equivalent were measured with a double-gridded ion chamber, henceforth referred to as a purity monitor, immersed in liquid argon. The purity monitor was based on the design in Reference~\cite{ref:purityMonitor}. A measure of electronegative impurities was determined from looking at the fraction of electrons generated at the cathode that arrived at the anode $(Q_A/Q_C)$ after a drift time, $t$. The fraction can also be interpreted as an electron lifetime, $\tau$ such that

\begin{equation}
\label{PrMElectronLifetime}
Q_A/Q_C = e^{-t/\tau}.
\end{equation}

A thorough description of the purity monitor and the data acquisition hardware and software can be found in Reference~\cite{ref:LAPD}. The purity monitor had a total drift distance from cathode to anode of 19~cm. It resided in a vessel downstream of the cryostat used for the breakdown tests. Before and after breakdown measurements were taken in the cryostat, argon was allowed to flow into the purity monitor vessel for a purity measurement. 

The systematic uncertainties of the purity measurement had the effect of decreasing the fraction of $Q_A/Q_C$, consequently decreasing the observed lifetime~\cite{ref:LAPD}. Therefore, the reported electron lifetimes from the purity monitor should be interpreted as a lower bound on the true lifetime.

\subsection{Experimental Procedure}

Measurements over the range of distances between 0 and 2.5~cm were collected over a 12 hour period for a given probe and contamination level. First, the zero spacing point was found using the force transducer described in Section \ref{subsec:HVC}, and the current was monitored as voltage was applied. The probe was then moved to a distance of 0.1 mm away from the ground plate and the voltage ramped up at a rate of 50~V/s until breakdown occurred. 
Breakdown voltage is defined as the voltage where the current drawn by the power supply changes its value by more than +100\% for a fraction of a second.

The cycle of ramping the voltage up until breakdown and recording the voltage at which the breakdown occurred was repeated multiple times at each distance. 
The distances considered in these tests are 0.0, 0.1, 0.3, 0.5, and 0.8 mm; 1 through 10 mm in millimeter steps; and 15, 20, and 25 mm.
The zero-point of the probe with respect to the ground plate was measured each time the probe was moved. For gap spacings of 1~mm and above, the HV ramp rate was increased from 50~V/s to 250~V/s. No significant effect on the breakdown voltages was found by making this change for the larger distances. Increasing the rate for the spacings above 1~mm allowed the full set of gap spacing measurements to be completed for a given test configuration before a significant amount of argon boiled off.

\begin{figure}[hbt]
  \centering
  \subfloat[]{\includegraphics[width=0.50\textwidth]{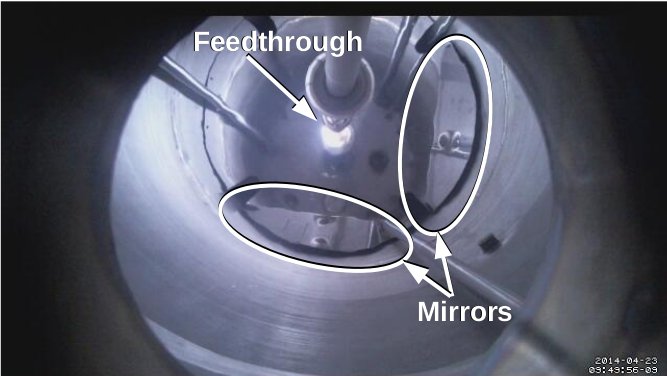}}
  \subfloat[]{\includegraphics[width=0.50\textwidth]{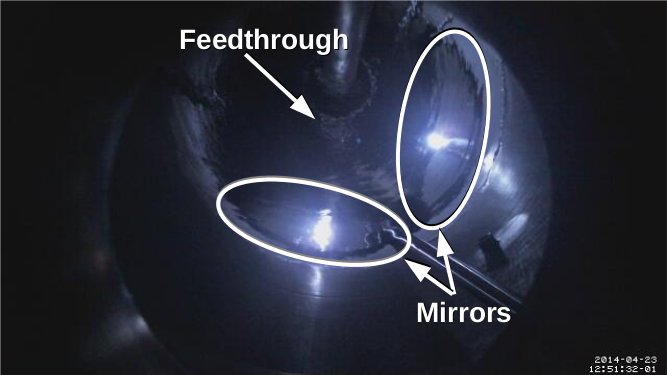}}
\caption{A view looking down at the mirrored grounding plate inside the liquid argon-filled test cryostat comparing (a) the effect of a spark along the feedthrough, versus (b) from the electrode to the grounded plate.
In the first case, spark light is only visible along the feedthrough and not in the mirrors. In the second case, light is present in the mirrors and not along the feedthrough.} 
\label{fig:pictures}
\end{figure}

With the geometry of the test setup, it was found that the liquid argon boil-off caused the liquid level to decrease to nearly the bottom edge of the ground tube at times during testing. In this situation, for gap spacings greater than 1~cm, breakdown would sometimes occur along the feedthrough instead of the sphere-plane gap as seen in Fig.~\ref{fig:pictures}. The breakdown was captured using  a camera mounted above the view ports. This effect was mitigated by increasing the liquid argon level in the cryostat. However, in order to avoid changing the contamination level of the liquid during a given test configuration, the liquid argon level was not increased until the end of data-collection for that configuration. In the case of the 5.0~mm probe, data for gap spacings larger than 1.5~cm were not considered because the liquid level was too low, and similarly in the case of the 76~mm probe, data for gap spacings larger than 1~cm were not considered.

\section{Results}

\begin{figure}[]
  \centering  
  \subfloat[]{\includegraphics[width=0.5\textwidth]{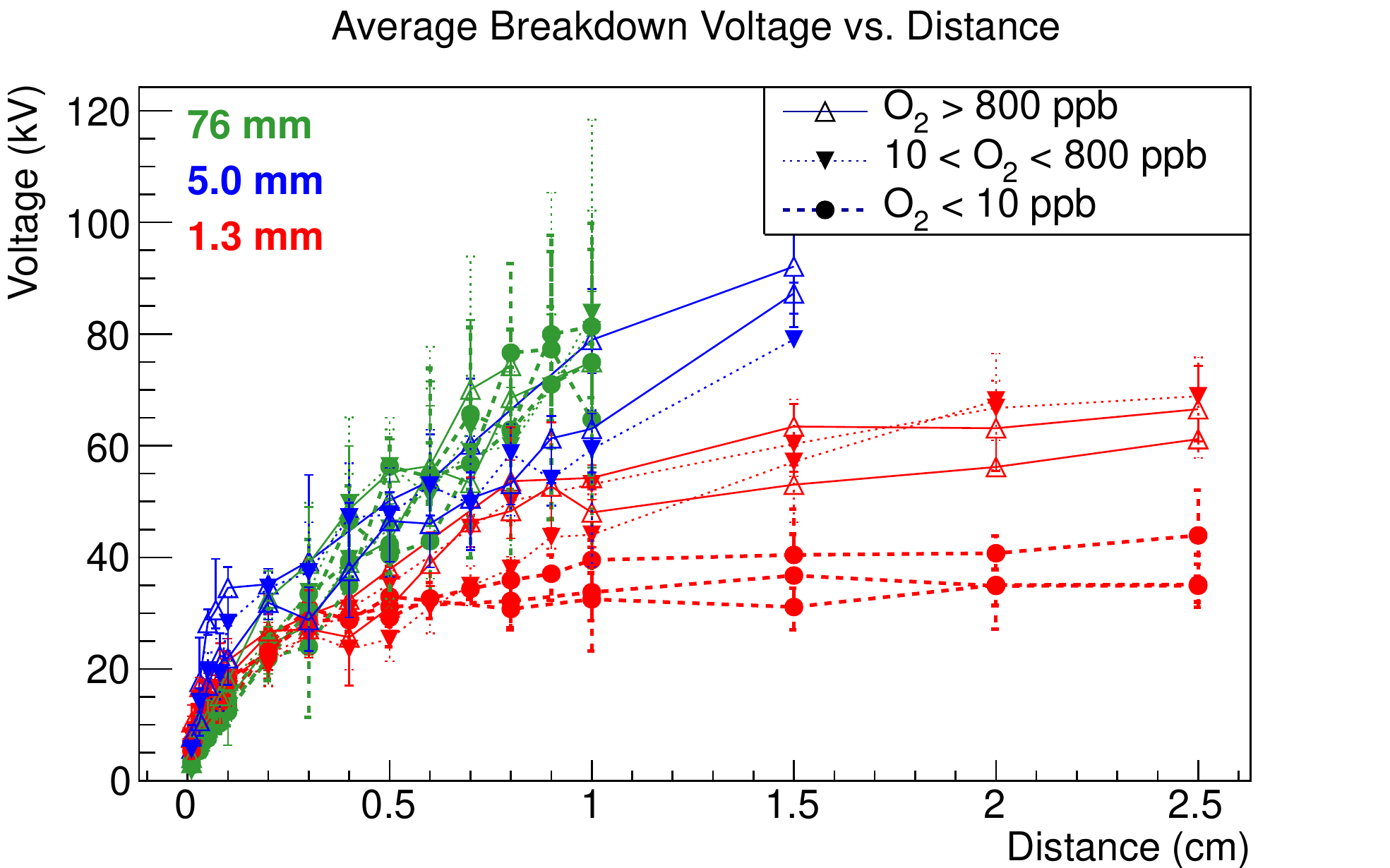}\label{fig:voltageVsDistance}}
\subfloat[]{\includegraphics[width=0.5\textwidth]{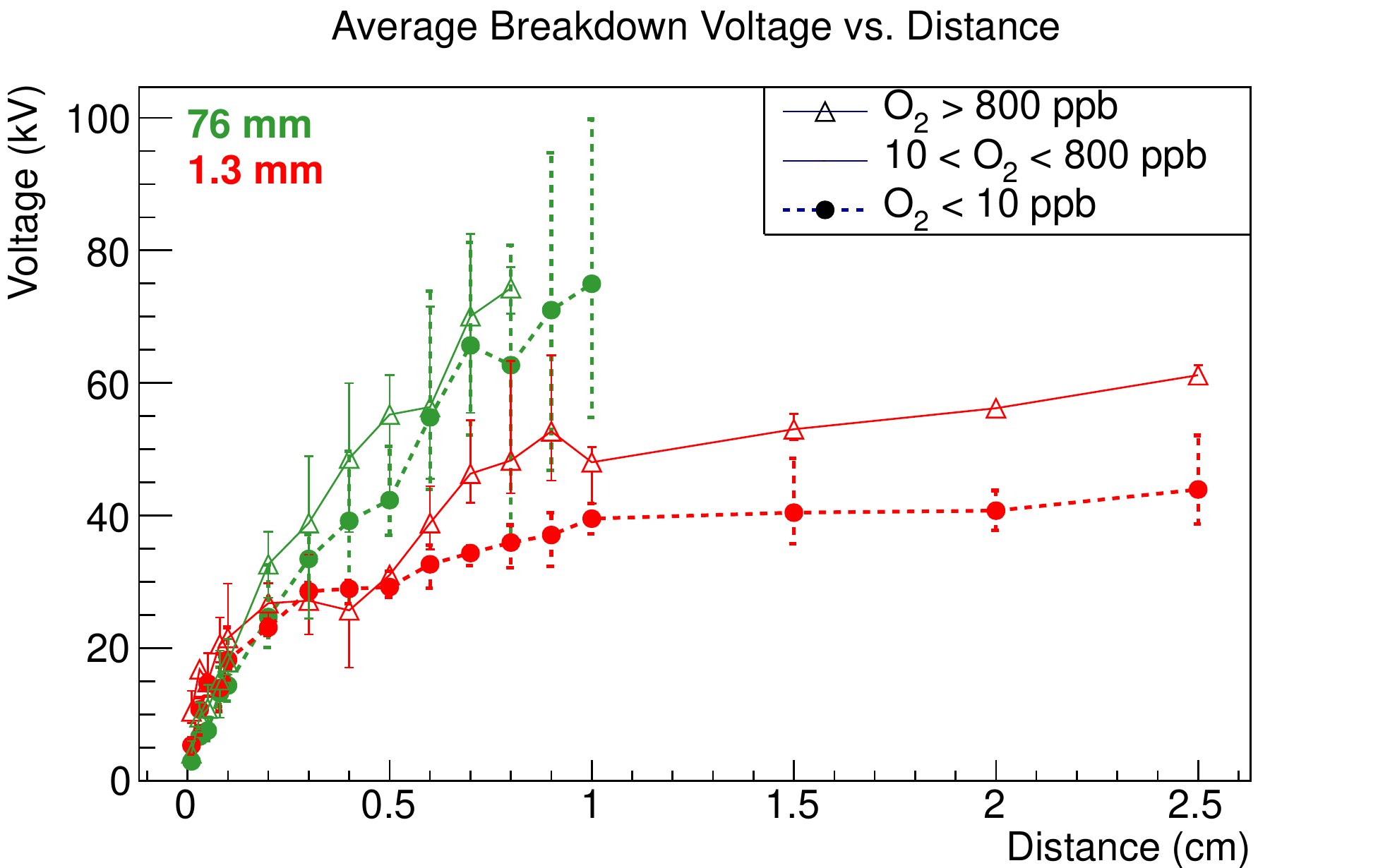}\label{fig:voltageVsDistance_extreme}}
  \caption{(a) The measured average breakdown voltage recorded as a function of distance for the three probes. 
  The 76~mm sphere data are shown in green, the 5.0~mm sphere data are in blue, and the 1.3~mm sphere data are in red.
  Different electronegative contamination levels are indicated by the solid, dashed, and dotted lines specified in the legend. (b) The highest and lowest electronegative contaminations levels data for the 1.3~mm and 76~mm spheres.}
  \label{fig:voltageVsDistances}
\end{figure}

A full list of the collected data, including probe size and contamination levels, is shown in Table~\ref{tab:dates}. Fig.~\ref{fig:voltageVsDistances} shows the average breakdown voltage versus distance for a given probe and electronegative contamination level. For each configuration, the average at each distance was computed from multiple breakdown voltages. 
The extrema of the uncertainties are the lowest and highest values for the given measurement.
Systematic uncertainties are not shown here since they would be dwarfed by the spread of the data. A 0.05~mm uncertainty was estimated for the gap spacing measurement based on the resolution of the digital linear readout.  A 500~V uncertainty was estimated on the breakdown voltage measurement based on the ramp rate and step size.

As can be seen from Fig.~\ref{fig:voltageVsDistances}, a dependence of the breakdown voltage on liquid argon electronegative contamination level is most evident with the 1.3~mm probe. 
At gap spacings of about 1~cm with the 1.3~mm probe, the breakdown voltages for oxygen contaminations between 0.2 and 1.4~ppm are about a factor of 1.5 more than for oxygen contaminations between 0.29 and 1.8~ppb.  No such dependence of the breakdown voltage on liquid argon electronegative contamination level is observed for the 76~mm probe in the same ranges of oxygen equivalent contamination. This difference in behavior can be seen in Fig.~\ref{fig:voltageVsDistance_extreme}, where the average breakdown voltage versus distance is plotted for only the highest and lowest electronegative contaminations levels achieved with the 1.3~mm and 76~mm probe.


\begin{table}{}
\centering
\begin{tabular}{c c c c}
\hline \hline{}
Probe (mm) & O$_2$ (ppb) & N$_2$ (ppm) & Lifetime (ms)\\ \hline 
1.3 &   $ 1200^{+    50}_{-    50}$ &  $1.3         ^{+0.1 }_{-  0.1}$  &  \\
&   $ 1200^{+    50}_{-    50}$ &  $1.7         ^{+0.2 }_{-  0.2}$  &  \\
&   $ 1200^{+    50}_{-    50}$ &  $1.7         ^{+0.2 }_{-  0.2}$  &  \\
&   $  744^{+    2}_{-  2}$ &  $3.1         ^{+0.2 }_{-  0.5}$  &  \\
&   $ 744 ^{+    2}_{-  2.0}$ &  $3.1 	    ^{+0.2 }_{-  0.5}$  &  \\
&   $ 1.8 ^{+    1.0}_{-  0.7}$ &  $6.0 	    ^{+0.2 }_{-  0.5}$  &  \\
&   $ 0.35^{+ 0.15}_{- 0.15}$ &  $5.2         ^{+0.2 }_{-  0.5}$  &  \\
&   $ 200 ^{+   20}_{-   10}$ &  $5.5         ^{+0.2 }_{-  0.5}$  &  \\
&   $ < 0.29^{\dag}         $ &  $6.4         ^{+0.2 }_{-  0.2}$  & $>1.0$  \\
&   $ < 0.29^{\dag}		$	&  $6.4         ^{+0.2}_{-0.2}$		& $ > 1.0 $ \\
&   $ 1400^{+  100}_{-  50}$ &  $28          ^{+2}_{-  2}$  &  \\
\hline
5.0 &   $ 1300^{+    50}_{-    50}$ &  $2.2         ^{+0.1 }_{-  0.1}$  &  \\
&   $ >900                  $ &  $22        ^{+5}_{-5}$  &  \\
&   $ 775^{+ 50}_{- 50}$ &  $22        ^{+5}_{- 5}$  &  \\
\hline
76 &   $ 1400^{+    50}_{-    50}$ &  $1.4         ^{+0.1 }_{-  0.1}$  & \\
&   $  70 ^{+  100}_{-   20}$ &  $3.2         ^{+0.2 }_{-  0.5}$  &  \\
&   $ 10.20^{+  0.05}_{-  3.00}$ &  $5.5         ^{+0.2 }_{-  0.5}$  &  \\
&   $ 0.60^{+ 0.10}_{- 0.10}$ &  $5.0         ^{+0.2 }_{-  0.5}$  &  \\
&  $ < 0.23^{\dag} $	        &   $5.9^{+0.2}_{-0.5}$		& $ > 1.3 $ \\
&   $ < 0.14^{\dag} $			&  	$5.9^{+0.2}_{-0.5}$		& $ > 2.1 $ \\
&   $ < 0.13^{\dag} $			&  	$5.9^{+0.2}_{-0.5}$		& $ > 2.3 $ \\
&   $ < 0.13^{\dag} $			&  	$5.9^{+0.2}_{-0.5}$		& $ > 2.3 $ \\
& $ < 0.09^{\dag}  $		&  	$5.9^{+0.2}_{-0.5}$		& $ > 3.5 $ \\
%
&   $ 1500^{+  100}_{-  50}$ &  $24          ^{+2 }_{- 2}$  &  \\
\hline \hline
\end{tabular}
  \caption{The configuration and contamination levels present during the breakdown tests. Errors on the contamination level are based on gas analyzer resolution and measurement time. Values marked with $^{\dag}$
  are upper limits for oxygen equivalent contamination as derived from the purity monitor
  measurements. Lifetime values listed are derived solely from the purity monitor.}
  \label{tab:dates}
\end{table}


\begin{figure}[]
  \centering
  \includegraphics[width=0.8\textwidth]{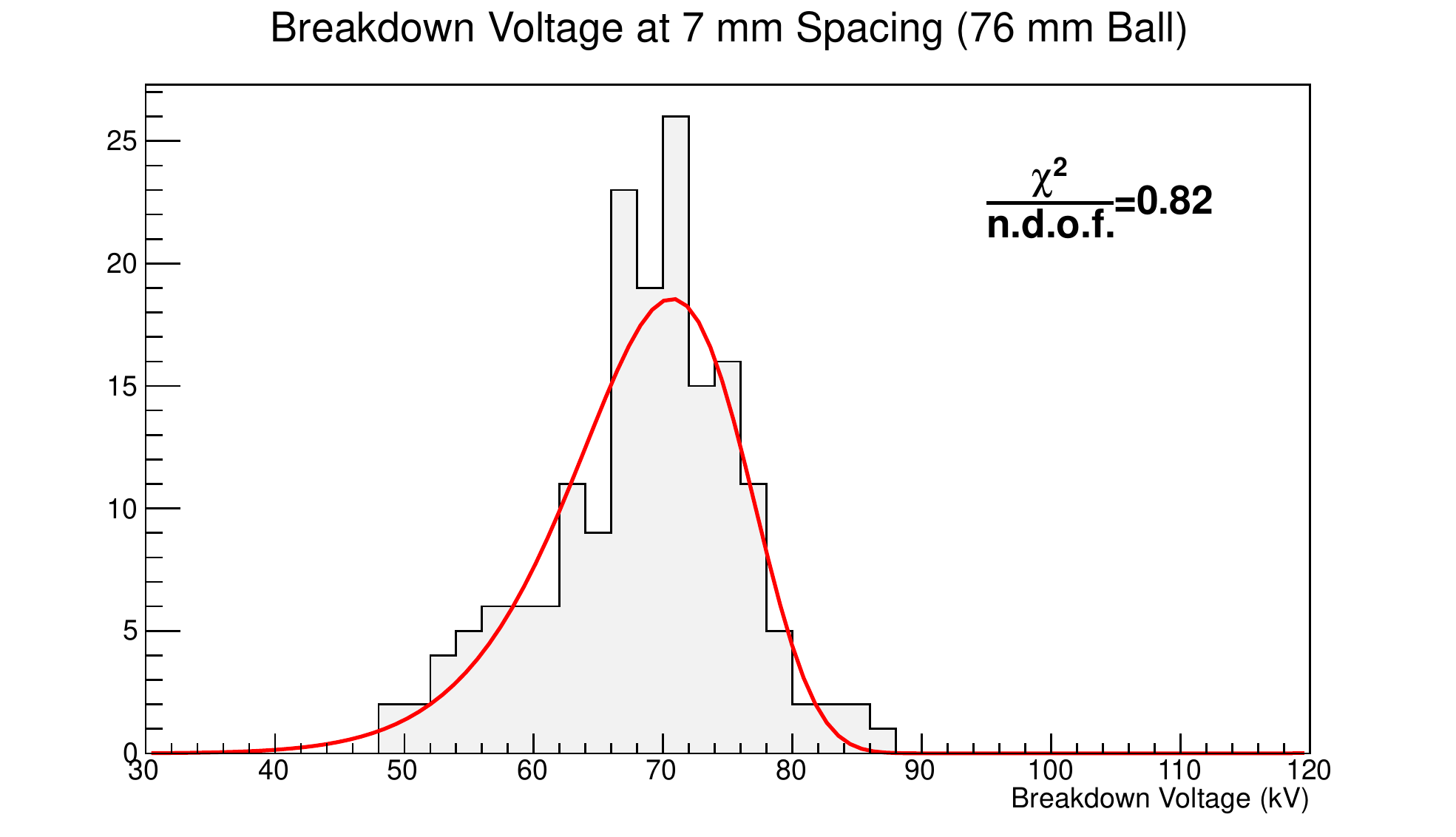}
  \caption{The distribution of breakdown values for the 76~mm ball for a fixed 7~mm gap spacing. A fit to the Weibull function is shown by the solid red curve.}
  \label{fig:rapidFireFit}
\end{figure}

In an attempt to understand the spread of the data, many breakdown voltages were recorded for a fixed distance and contamination level with a given probe. In Fig.~\ref{fig:rapidFireFit}, 174 breakdown voltages are plotted for the 76~mm probe with a 7~mm gap spacing. The data in Fig.~\ref{fig:rapidFireFit} are shown fit with a Weibull function~\cite{ref:gauster}, which demonstrates good agreement. 
The underlying mechanism that produces an asymmetric distribution of voltages is not understood. Previous measurements in other liquid insulators \cite{ref:gerhold2, ref:gerhold, ref:toil} observed similar distributions that were also well described by a Weibull function, suggesting that the process of electrical breakdown has a stochastic nature in liquid insulators. 

\begin{figure}[]
  \centering
  \includegraphics[width=0.8\textwidth,angle=270,trim=0in 0.5in 0in 2in,clip]{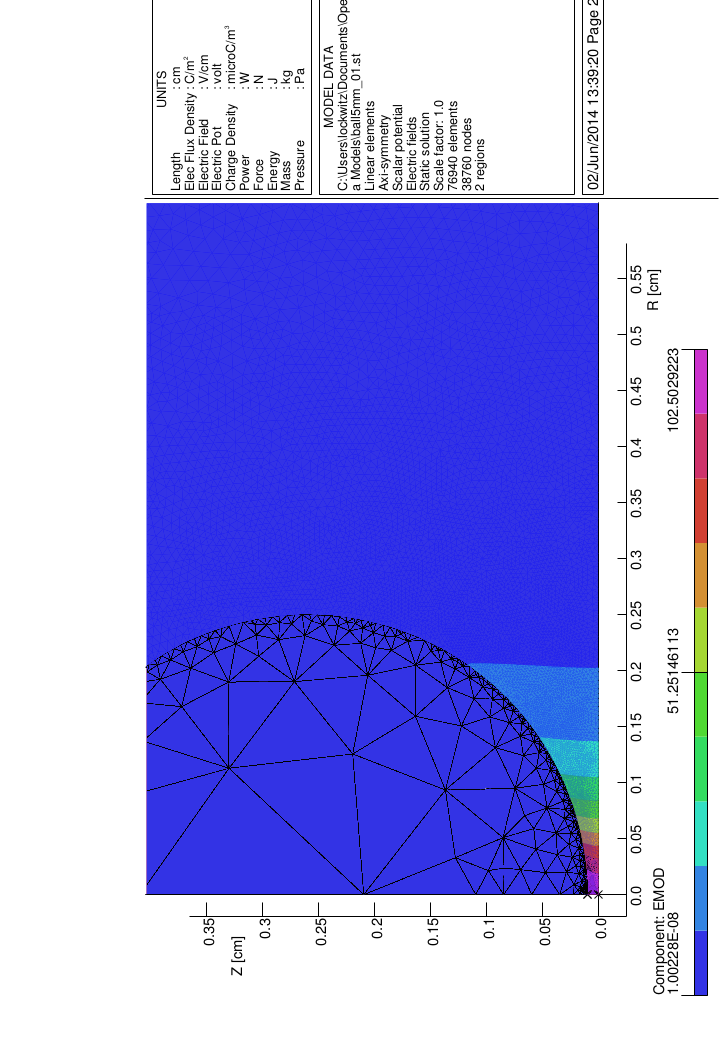}
  \caption{Sample electric field from {\sc Opera-2d}. This is for the 5.0~mm diameter probe at a 0.01~cm gap spacing.}
  \label{fig:operaEField}
\end{figure}

\begin{figure}[]
  \centering
  \includegraphics[width=0.8\textwidth]{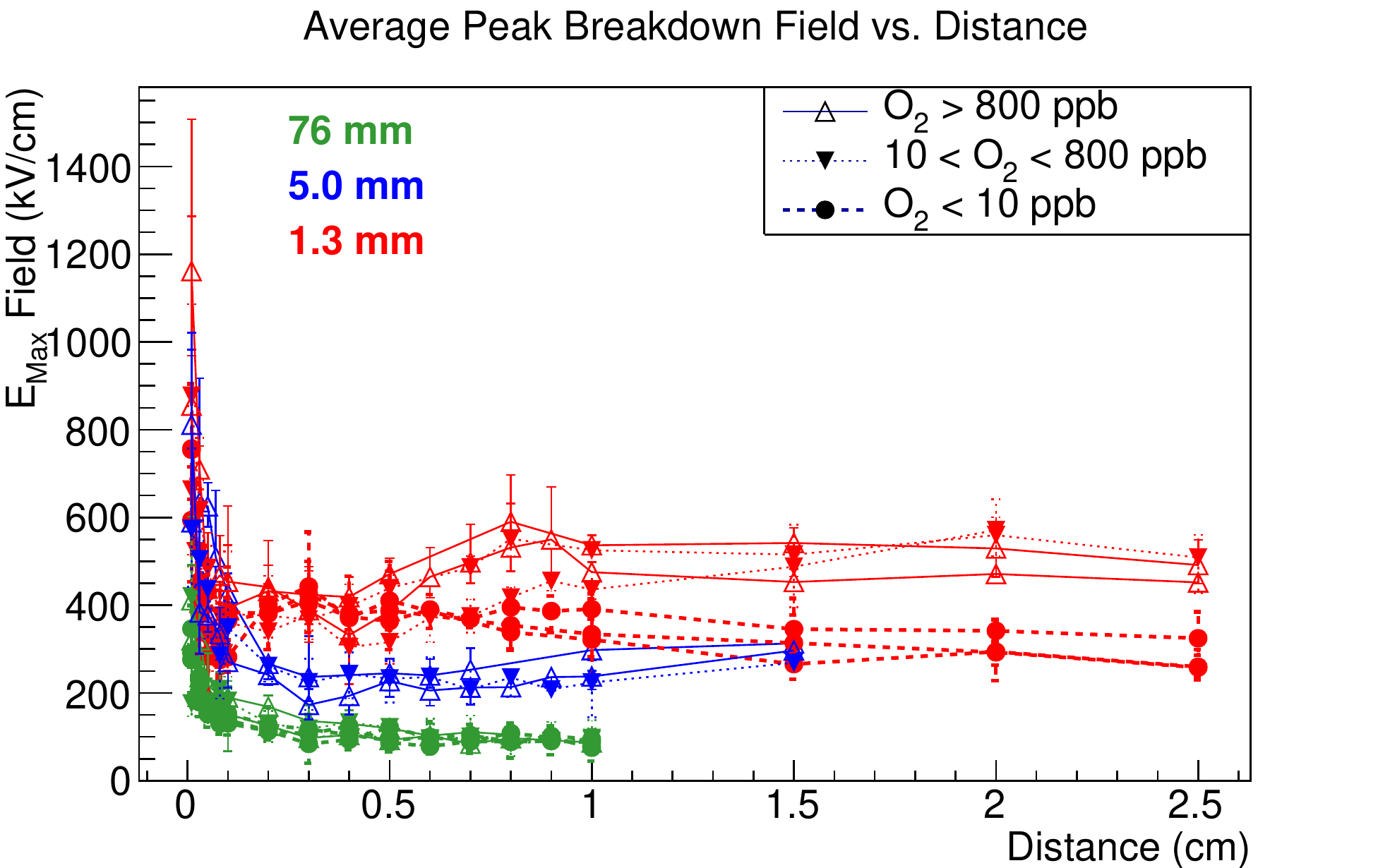}
  \caption{The average maximum breakdown field as a function of distance. The 76~mm sphere data are shown in green, the 5.0~mm sphere data are in blue, and the 1.3~mm sphere data are in red. The solid, dashed, and dotted lines are indications of the different electronegative contamination levels specified in the legend.}
  \label{fig:eFieldVsDistance}
\end{figure}

For each configuration, the electric field was computed using {\sc Opera-2d}. 
The field for the 5.0~mm probe at a spacing of 0.1~mm is shown as an example in Fig.~\ref{fig:operaEField}.
The average breakdown voltage values from Fig.~\ref{fig:voltageVsDistances} were translated to maximum electric field values using the OPERA model and the results are shown as average maximum breakdown fields in Fig.~\ref{fig:eFieldVsDistance}.

Gerhold \emph{et al.} suggest a minimum electric field value, below which, no breakdown will occur~\cite{ref:gerhold}.
The lowest breakdown field observed out of all probes and distances examined was 38~kV/cm. The value of 38~kV/cm can then be interpreted as an upper bound on the minimum electric field for breakdown to occur. 
Beyond providing a possible lower threshold, the maximum electric field is likely not a parameter of interest by itself, given the differing values among the probes. 

Gerhold {\emph {et al.}~\cite{ref:gerhold} additionally suggest a dependence between the maximum electric field at breakdown and a stressed cathode area in liquid helium data. The stressed area is defined as the area on the cathode surface with electric field exceeding some percentage, $\epsilon$, of the maximum electric field. In Figs.~\ref{fig:area80} and \ref{fig:area90}, the average maximum electric field is plotted versus stressed area for $\epsilon$ values 80\% and 90\% respectively. 
This representation unifies the breakdown behavior across the geometries tested in this study.
A trend in the data is observed, which appears at any contamination level. An example is given in Fig.~\ref{fig:areaac}, where the plots of Fig.~\ref{fig:areaa} are reported only for contaminations above 800~ppb oxygen equivalent.  
The cluster of points seen for the largest areas of the 1.3~mm and 5~mm probes is a result of the stressed area covering the entire surface of the probe's bottom hemisphere; any further increase in gap spacing does not significantly increase the stressed area.

The combined data for the three probes appears to be consistent with the function suggested in~\cite{ref:gerhold}:
\begin{equation}
\label{eq:aaa}
E_{max} = C(A)^{-0.26}
\end{equation}
\noindent
where $E_{max}$ is the maximum electric field at breakdown, $A$ the stressed area, and $C$ a constant term. The red line shown in Figs.~\ref{fig:areaa} and \ref{fig:areaac} is a representation of Eq.~\ref{eq:aaa}.

\begin{figure}[]
  \centering
  \subfloat[$\epsilon=0.8$]{\includegraphics[width=0.5\textwidth]{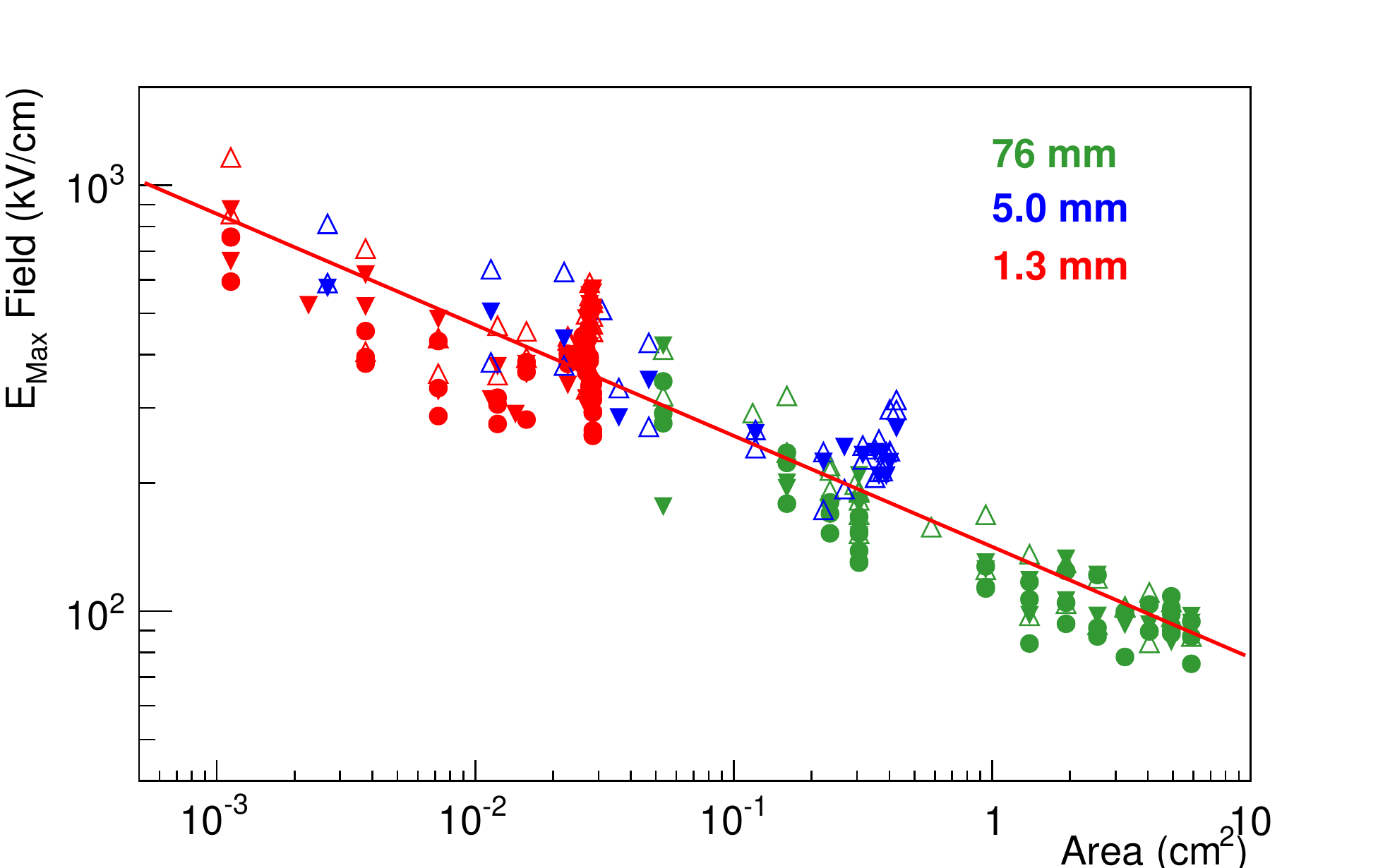}	\label{fig:area80}}
  \subfloat[$\epsilon=0.9$]{\includegraphics[width=0.5\textwidth]{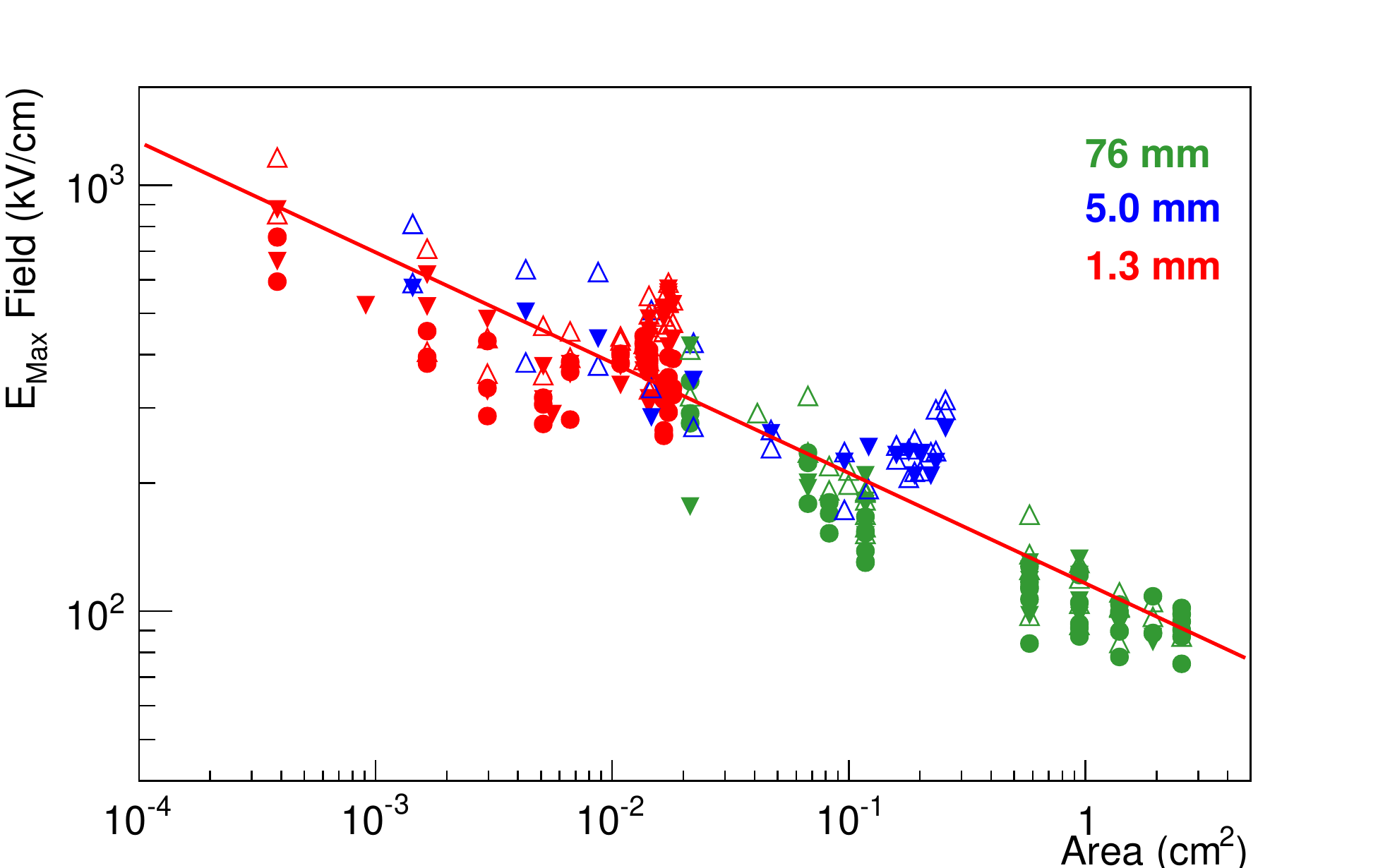}\label{fig:area90}}
\caption{
The average maximum breakdown field versus stressed area of the cathode. (a) The stressed area is defined as the area with an electric field greater than 
80\% of the maximum electric field. (b) The stressed area is defined as the area with an electric field greater than 90\% of the maximum electric field. 
} 
\label{fig:areaa}
\end{figure}

\begin{figure}[]
  \centering
  \subfloat[$\epsilon=0.8$]{\includegraphics[width=0.5\textwidth]{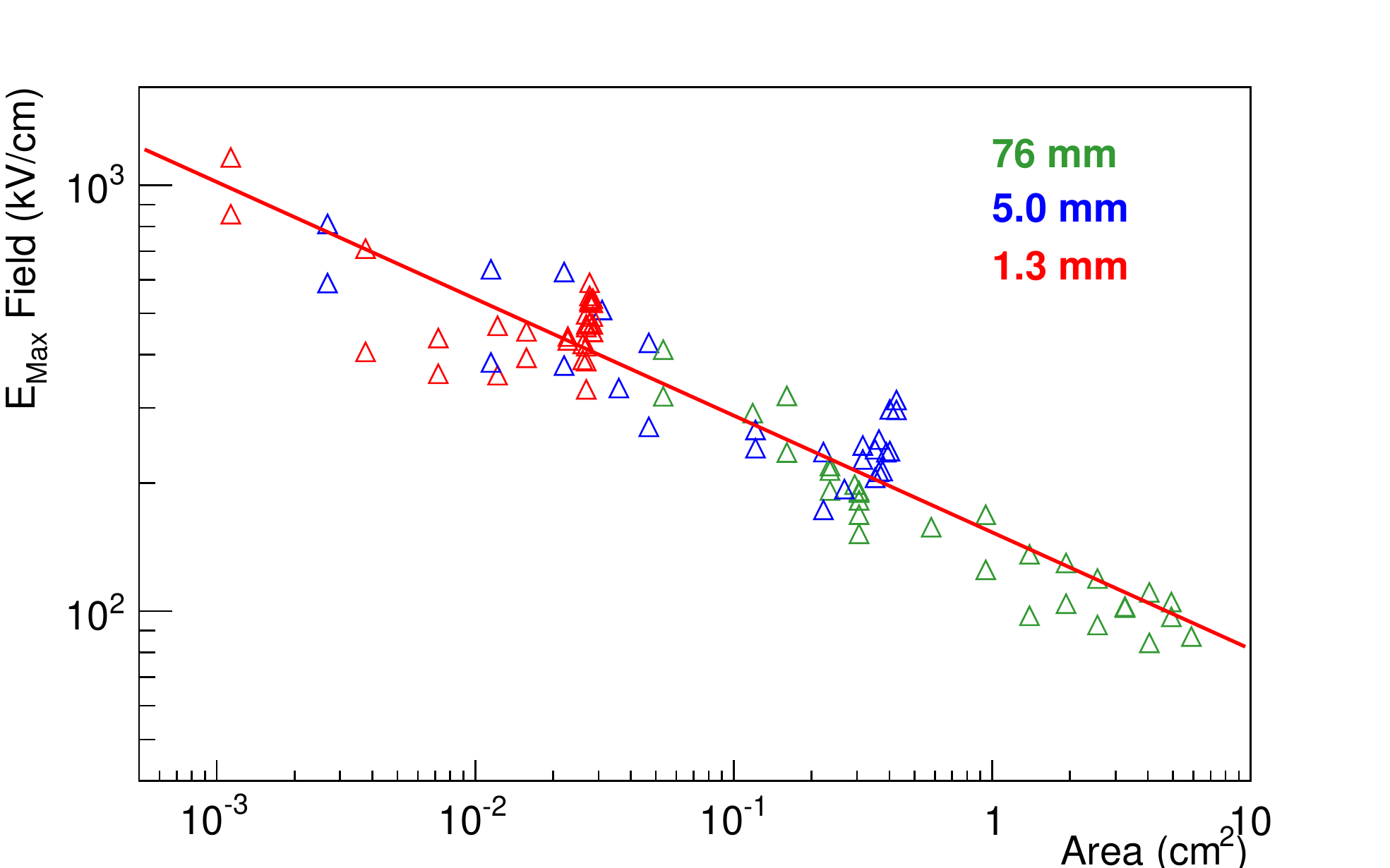}\label{fig:areaC80}}
  \subfloat[$\epsilon=0.9$]{\includegraphics[width=0.5\textwidth]{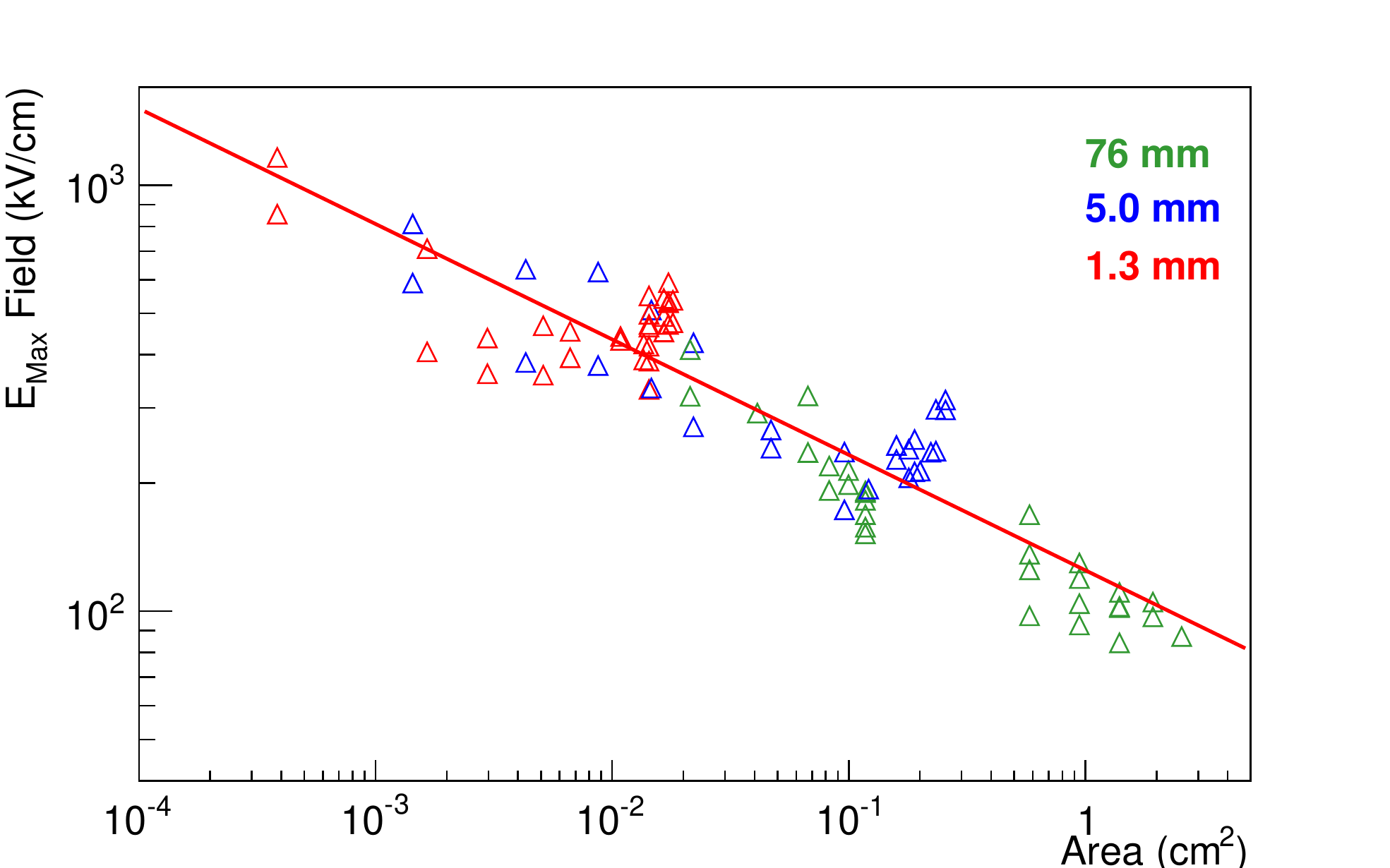}\label{fig:areaC90}}
\caption{The average maximum breakdown field versus stressed area of the cathode for contaminations above 800 ppb oxygen. (a) The stressed area is defined as area with an electric field greater than 80\% of the maximum electric field. (b) The stressed area is defined as area with an electric field greater than 90\% of the maximum electric field. } 
\label{fig:areaac}
\end{figure}

Gerhold {\emph {et al.}~\cite{ref:gerhold} further suggest a dependence of breakdown maximum electric field versus stressed volume. Preliminary calculations show a similar trend as the area effect.  This is expected since stressed volume and area are highly correlated in this geometrical setup, however a different geometry could be designed to better decouple the two parameters in a future test setup.

\section{Comparison to Previous Measurements}

The data presented in this paper affirm the growing consensus in the literature that breakdown voltages in liquid argon present a strong dependence on the geometry of the experimental setup and a weaker dependence on argon electronegative contamination level. 
Fig.~\ref{fig:bern} shows the data from this study along with two others that span a range of parameter space in terms of electrode distances and contamination levels.

\begin{figure}[]
  \centering
 \includegraphics[width=1.0\textwidth]{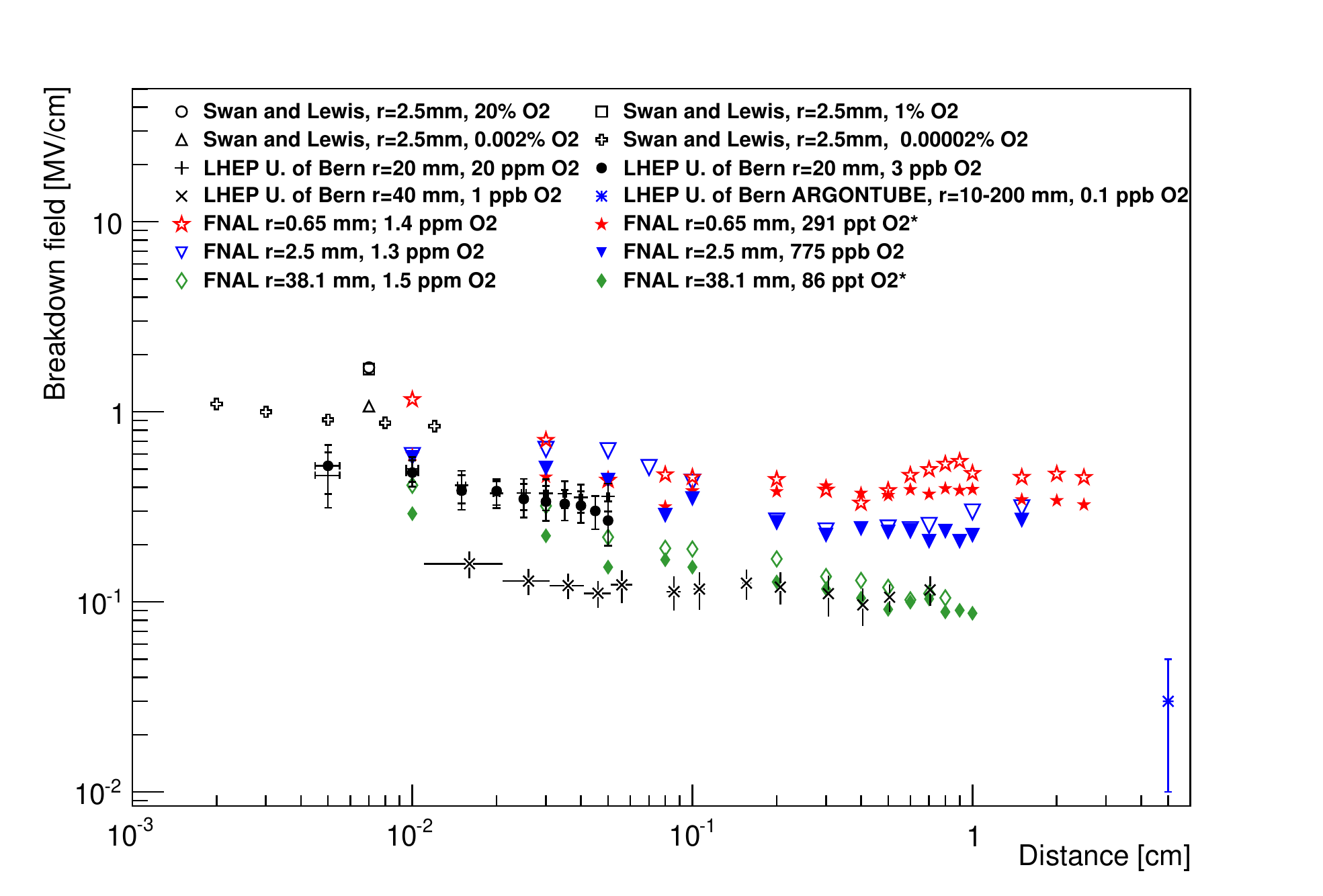}
\caption{A comparison to previous measurements~\cite{ref:bern}. The measurements described in this paper are shown in red, blue, and green and result from the highest and lowest contamination level measurement taken with each probe.  
The oxygen values with an asterisk are extrapolated from purity monitor measurements.}
\label{fig:bern}
\end{figure}
Early studies~\cite{ref:swan,ref:swan2} first indicated a possible influence of electronegative contaminants on breakdown voltages, but not for contamination levels or distances important for understanding LArTPCs.
The results from these early studies indicated that at oxygen levels above 0.2~ppm, the breakdown electric fields were above 1~MV/cm at sub-mm spacings for a sphere-sphere geometry. 
Those same studies showed that the breakdown electric field decreased with lower levels of oxygen.
These older data have often been cited as justification for the claim that liquid argon could sustain electric fields on the order of 1~MV/cm. 
Recent studies using the same sphere-plate geometry as reported in this paper~\cite{ref:bern} found that argon electronegative contamination level had a relatively small impact in comparison to geometry in influencing the breakdown field.

\section{Conclusion}

A study of the dielectric strength of liquid argon has been performed wherein both the argon electronegative contamination level and
the electrode size and separation have been controlled. 
The results show a dependence
of the electric field strength at breakdown on the geometry of the test. Breakdown occurs at lower field strengths as the separation
between electrodes increases from sub-millimeter to centimeter scales. 

Breakdown also occurs at lower field strength as the stressed surface area of the electrode increases; the maximum electric field at breakdown scales roughly as the stressed area to the $-0.3$ power. 
This study is the first time that the dependence of the breakdown field on stressed cathode area has been shown for liquid argon. 

There is some dependence between breakdown field and
argon electronegative contamination level, but the effect from contamination level is small relative to that of the geometries used in this study. 
The 1.3~mm electrode experienced breakdown at lower electric fields when the argon had lower contamination level: breakdown electric field values for oxygen contaminations between 0.2 and 1.4~ppm were about a factor of 1.5 more than for oxygen contaminations between 0.29 and 1.8~ppb at gap spacings of about 1~cm.
However, the 76~mm electrode showed no clear dependence between electric field at breakdown and argon electronegative contamination level.  

Electrical breakdown in liquid argon is a stochastic process; the chain of events which lead to the breakdown
is not precisely repeatable. A given geometric setup experienced breakdown over a range of voltages; marks left by sparks on the surface of the larger electrodes show that the location of the breakdown can shift when all else is held static.
This observation is consistent with the hypothesis mentioned in \cite{ref:bern} that space charge effects in the stressed volume near the cathode contribute to breakdown probability.

The observed geometric effects are of critical
importance in the design of LArTPC detectors for neutrino experiments, where drift distances must be on the scale of meters.
Ideally, one would use a well-known breakdown probability distribution to design LArTPCs. A suitable operating point could then be chosen to minimize damage or dead time while accounting for signal-to-noise requirements. It is difficult in practice to extrapolate these measurements to TPCs given their unique and complicated geometries.  
However, it is possible that stressed area may need to be minimized in TPC design to prevent dielectric breakdown.

\section{Acknowledgements}

The authors wish to thank the staff at Fermilab for their technical assistance, specifically Jim Walton. Additionally, a great deal of thanks is due to the authors of~\cite{ref:bern} from the Laboratory for High Energy Physics at University Bern for graciously providing their data. 

Fermilab is operated by Fermi Research Alliance, LLC under Contract No. De-AC02-07CH11359 with the United States Department of Energy.

\bibliographystyle{utphys}
\bibliography{references}

\end{document}